# Enhancing Source Code Representations for Deep Learning with Static Analysis

Xueting Guan
School of Computing and Information Systems
The University of Melbourne
Melbourne, Australia
guaxg@student.unimelb.edu.au

Christoph Treude
School of Computing and Information Systems
The University of Melbourne
Melbourne, Australia
christoph.treude@unimelb.edu.au

## ABSTRACT
Deep learning techniques applied to program analysis tasks such as code classification, summarization, and bug detection have seen widespread interest. Traditional approaches, however, treat programming source code as natural language text, which may neglect significant structural or semantic details. Additionally, most current methods of representing source code focus solely on the code, without considering beneficial additional context. This paper explores the integration of static analysis and additional context such as bug reports and design patterns into source code representations for deep learning models. We use the Abstract Syntax Tree-based Neural Network (ASTNN) method and augment it with additional context information obtained from bug reports and design patterns, creating an enriched source code representation that significantly enhances the performance of common software engineering tasks such as code classification and code clone detection. Utilizing existing open-source code data, our approach improves the representation and processing of source code, thereby improving task performance.

## 1 Introduction

The increase in software applications has made understanding code more challenging, requiring tools to aid in this process. Contextual factors like control flow, version updates, and bug reports have been shown to enhance code understanding [1]. For example, a developer perception model showed the importance of these contextual elements in understanding tasks [6]. Simultaneously, open-source software communities have provided extensive code datasets, while advancements in artificial intelligence drive intelligent software evolution [12].

In natural language processing, models such as BERT have shown strong results [5]. Following their success, similar models have been adapted to represent programming source code, aiding tasks like code summarization. However, these models often overlook the inherent structural and semantic nuances of source code [2]. Recent studies have attempted to enhance code representation models by using advances from deep learning [13]. Yet, these methods frequently disregard crucial information contained in documentation and bug reports, which could elucidate potential errors [8]. Investigating source code structures, like design patterns, might offer a solution [28].

Our research assesses the effectiveness of enriching the ASTNN method [2]—a method that capitalizes on AST-based neural networks to capture sentence-level lexical and syntactic knowledge—with additional context about bugs and design patterns. Since the ASTNN implementation is publicly accessible, we can replicate the approach and compare our method effectively. This enrichment is evaluated within the scope of prevalent software engineering tasks [10]. Our goal is to use models that can process both additional information and code representation at the same time to improve efficiency in subsequent tasks include clone detection and code classification. Although the integration of bug report data did not notably augment the performance, we discovered that the inclusion of design pattern information was instrumental in enhancing the precision of subsequent tasks.

While integrating contextual information into source code representations has proven beneficial, the type of context and its integration method can greatly influence the performance of downstream tasks [3]. This paper provides an overview of relevant work, presents a preliminary study, and discusses research findings, its limitations, and potential areas for future research.

## 2 Related Work

Our work contributes to ongoing research at the nexus of source code representation and static analysis in the realm of software engineering.

### 2.1 Source Code Representation

Recent strides in machine learning have revolutionized source code representation, spawning methods classified into four tiers: text-based, lexical, syntactic, and semantic [9]. Each level carries distinct characteristics, along with associated advantages and drawbacks. For instance, text-based representations such as n-gram models, trained on expansive corpora, prove effective in predicting tokens across various domains [16]. Lexical-based representations offer abstraction capabilities, while syntactic and semantic-based representations afford higher abstraction levels but demand preprocessing, such as converting source code into tree or graph structures [17]. Syntactic and semantic-based methods frequently harness Abstract Syntax Trees (AST), like ASTNN, which learns syntactic knowledge from smaller subtrees. Semantic-based methods often weave in code dependency information relating to data and control flow. The integration of low-level syntactic and high-level semantic information serves to enhance source code representation for program comprehension tasks.

The value of adding context to deep learning models has been underscored in recent research, especially encoding call hierarchy context with code information [3]. Besides, earlier methods overlook the specific context essential for effective source code representation [21]. Our work aims to incorporate static analysis of code, which could potentially boost the performance of various methods employed in source code representation.

## 2.2 Bug Information

The significance of context is well-recognized in software engineering. Bug reports, for example, are critical to bug triage and serve as an invaluable source for code representation. The quality of information contained in bug reports can substantially influence process efficiency and accuracy—imprecise reporting often results in unreliable outcomes [20]. Studies have uncovered frequent occurrences of incorrect or incomplete information in bug reports, as well as misclassification issues like non-bugs being erroneously labelled as bugs [18]. The integration of accurate bug information with code representation could affect subsequent tasks. For this paper, we use the term 'bug' to denote bug information supplied by automated tools, which identify existing latent bugs in source code.

## 2.3 Design Patterns

Design patterns hold a prominent position in software development, offering invaluable insights into code functionality. Existing methods for automated design pattern recognition span across multiple approaches, including graph theory-based, formal technique-based, software metrics-based and artificial intelligence-based [27]. Our study explores factors that influence the preparation and optimization of training samples in design pattern recognition, and how integrating design pattern information can enhance ASTNN representation performance in subsequent tasks.

We posit that the amalgamation of bug reports and design-pattern-based static analysis could significantly enrich the input of source code representation methods. Much like human developers, deep learning models may benefit from the insights gained through static analysis. In our study, we leverage existing models, infuse pertinent static analysis information, and examine their influence on the performance of source code representation.

## 3 Preliminary Study

This section explores enhancing deep learning models for software engineering tasks through static analysis. The inclusion of bug and design pattern information as contextual data leads us to hypothesize that such an approach can be generalized across different code datasets and diversified software engineering tasks.

### 3.1 Research Questions

To probe the viability of our approach and determine the most effective implementation, we formulated two research questions:

**RQ1:** What is the impact of integrating bug report information on the performance of ASTNN model?

**RQ2:** How does the inclusion of design pattern information influence the performance of ASTNN model?

### 3.2 Data Collection

Our study relies on two public datasets, mirroring the baselines for code classification and clone detection tasks that were used in the original ASTNN work [2]. These datasets enable subsequent comparisons. Additionally, we used supplementary datasets for model selection in bug report data and design pattern detection.

*Dataset for Code-Related Tasks.* We adopted two datasets as benchmarks for subsequent tasks. The first dataset comprises 104 programs all implementing the same function [4]. The second dataset, BigCloneBench (BCB) [14], specifically caters to clone detection methods by focusing on syntactic similarity between codes. These established datasets provide a foundational layer, augmented with additional static analysis information such as bug reports and design pattern data using specific tools and methods. We adhered to the experimental settings provided by previous work to ensure a thorough comparison [2].

*Dataset for Bug Detection.* We enriched the original dataset with bug information using FindBugs, a Java bug detection plugin for Eclipse. Its superior performance and efficiency over other bug detection tools have been attested in previous studies [26]. The FindBugs bug dataset acted as a baseline for exploring enhanced defect detection classifiers to identify valuable bug information. About 50% of the packages showed no observable bugs, while the rest had up to 103 defect reports in this research.

*Dataset for Design Pattern Detection.* Our study used a labeled Java corpus as the base corpus, expanded by incorporating other publicly available code data like the GitHub Java corpus (GJC) [11] and the DPDF corpus [27]. We used this extended labeled code dataset from GJC to explore an effective design pattern recognition approach that could be applied to the original dataset.

### 3.3 Bug Detection

The FindBugs Eclipse Plugin was deployed to enhance the original dataset with bug information. Custom bug detectors embedded within the source code augmented the process's effectiveness. We followed validated recommendations from prior research [26] to configure the tool's "medium" warning settings. Our bug detection aimed to improve dataset precision, considering some bug reports might lack substantial information. The process for the automatic bugs classification began with preprocessing and parsing the dataset. Then, the N-gram IDF was applied to the pre-processed corpus of bug reports, generating a list of valid N-gram key terms. For each bug report, the raw frequency of each N-gram item was computed and stored as membership vectors. These vectors were then combined with an existing dataset from previous work which contains the correct bug report type for each bug report file [26]. The resulting combined vectors served as input to train the classification model using logistic regression and random forest as classification models. Once the model was effectively trained, it could filter out misinformation as accurately as possible.

By leveraging the FindBugs tool and optimized models, our foundational dataset was enriched with meaningful bug information. As illustrated in Figure 1, a code fragment is presented along with its associated bug information. The dataset, now furnished with bug information in textual description, was processed subsequent phases and then fed into the ASTNN model.

```
protected static boolean allEmpty(Name name){
    Enumeration enum_ = name.getAll();
    while (enum_.hasMoreElements()){
        if (!enum_.equals("")){
            return false;
        }
    }
    return true;
}
// ------ call to equals() comparing unrelated class and interface
```

**Figure 1: A code fragment with the bug information**

### 3.4 Design Pattern Detection

Design pattern detection relies on feature extraction to navigate the complexity of programming languages and draw out semantically significant concepts [7]. These features capture the intercommunication pathways and organizational structures within the code. Based on previous research, we segmented the selection of features into two categories: class-level features and method-level features [27]. To address the classification challenge, we constructed multi-class classifiers using the Scikit-learn library in Python. For setting the learning parameters of the classifier, the number of estimates was set to 100, the learning rate to 1, and the AdaBoost algorithm was applied. To ensure unbiased predictions, Layered K-Fold Cross-Validation was employed to evaluate the performance of the machine learning models on code data samples.

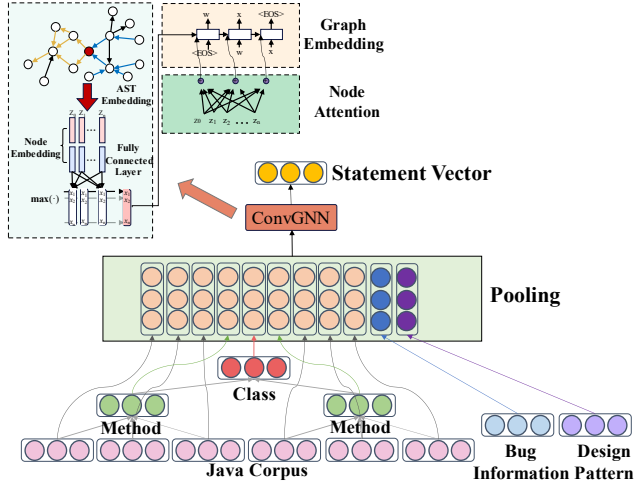

Figure 2: The structure of statement encoder

Various machine learning techniques can be applied to integrate and process data of different types. In our work, as illustrated in the Figure 2, max-pooling is employed to combine source code with supplementary bug information and design pattern information. Then, encoding is performed using a GNN-based encoder. Consequently, the resultant statement vector can be input into the ASTNN model to generate a representation of the source code, which proves instrumental for subsequent tasks [3].

The experimental results are shown in Table I and Table II.

### 3.5 Results

For the code classification task, we evaluated the classifier's effectiveness using the accuracy rate on the test set. Given that code clone detection can be viewed as a binary classification problem, determining the presence or absence of cloning, we deemed evaluation indicators from previous studies, such as the F1 score and accuracy, suitable for comparing the classifier's performance.

The results of the code classification task, based on the OJ dataset and additional datasets, are depicted in Table I. The results suggest that without any incorporation of static analysis information, the ASTNN model achieved an accuracy score of 96.6%. The introduction of Eclipse bug information slightly degraded the performance, resulting in an accuracy score of 95.6%. However, when integrating additional information from filtered bug reports, the performance improved, yielding an accuracy score of 97.8% (a 2% improvement). These outcomes suggest that superfluous information could adversely affect performance, reinforcing the necessity for efficient filtering to optimize results.

The results of the code clone detection task on the BCB dataset are displayed in Table II. The dataset was partitioned into five segments, each corresponding to a different sample type, and each type was subjected to detection. The results revealed that without the inclusion of static analysis, the ASTNN model yielded an F1 score of 84.9% on the BCB-ALL data. However, the inclusion of static analysis substantially improved the performance, with F1 scores of 90.8% and 93.2% achieved for the addition of bug reports and design patterns, respectively. The best performance was obtained when both bug reports and design patterns were added, resulting in an F1 score of 94.6%, a 9.7% improvement. These findings indicate that the combined use of static analysis information is more effective than its individual use. And our research questions yielded the following answers.

**RQ1:** The goal of this research question was to determine the influence of integrating bug report information on the performance of ASTNN-based models. Our results indicate an improvement in accuracy when filtered bug report data was incorporated into the model. Intriguingly, when the bug information provided by the FindBugs tool was used, the accuracy rate was slightly lower than the original performance. This outcome suggests that added information may interfere if it is redundant. To address this issue, a machine learning-based bug detection method was employed to filter the generated bug information. This process eliminated about 30% of the bug information, yielding a more meaningful bug information and a slight improvement in performance. The results suggest that while bug information impacts the code classification task, the effect is not profound. This could be due to the ASTNN model's robust performance on the original dataset.

Considering the code classification task results, we conducted an analysis to ascertain why the bug information had the observed effect on performance. We observed that the primary purpose of the code classification task is to classify the functions of code fragments. However, bug information primarily pertains to grammatical or structural errors in the code, which do not directly reflect the function of the code fragment. This realization prompted us to consider incorporating other types of information to enhance performance. Design patterns, as discussed in the next question, emerged as a source of functional information.

TABLE I. PERFORMANCE IN THE CODE CLASSIFICATION

| Metric (%) | Original dataset | FindBugs tool | Bug filter | Design pattern | Bugs filter + Design pattern |
|---|---|---|---|---|---|
| Accuracy | 96.6 | 95.6 | 97.8 | 98.2 | 98.5 |

TABLE II. PERFORMANCE IN THE CODE CLONE DETECTION

| | Original dataset | | | | FindBugs tool | | | | Bug filter | | | | Design pattern | | | | Bugs filter + Design pattern | | | |
|---|---|---|---|---|---|---|---|---|---|---|---|---|---|---|---|---|---|---|---|---|
| Metric (%) | A | P | R | F1 | A | P | R | F1 | A | P | R | F1 | A | P | R | F1 | A | P | R | F1 |
| BCB-T1 | 100 | 100 | 100 | 100 | 100 | 100 | 100 | 100 | 100 | 100 | 100 | 100 | 100 | 100 | 100 | 100 | 100 | 100 | 100 | 100 |
| BCB-T2 | 100 | 100 | 100 | 100 | 100 | 100 | 100 | 100 | 100 | 100 | 100 | 100 | 100 | 100 | 100 | 100 | 100 | 100 | 100 | 100 |
| BCB-ST3 | 98.7 | 96.3 | 95.2 | 95.7 | 94.2 | 93.6 | 92.1 | 92.8 | 9.7 | 98.7 | 96.7 | 97.7 | 99.2 | 99.1 | 98.6 | 98.8 | 99.9 | 99.3 | 98.7 | 99.0 |
| BCB-MT3 | 93.8 | 93.4 | 92.1 | 92.7 | 87.6 | 85.9 | 84.7 | 85.3 | 95.8 | 93.8 | 92.6 | 93.1 | 96.2 | 93.8 | 94.7 | 94.2 | 98.5 | 97.8 | 97.1 | 97.4 |
| BCB-T4 | 90.1 | 83.2 | 81.7 | 82.4 | 85.9 | 81.3 | 77.8 | 79.5 | 93.8 | 90.6 | 86.4 | 88.5 | 92.7 | 91.4 | 92.6 | 91.9 | 94.8 | 91.7 | 92.8 | 92.2 |
| BCB-ALL | 92.6 | 87.7 | 82.4 | **84.9** | 88.6 | 82.7 | 80.9 | **81.8** | 94.2 | 91.8 | 89.9 | **90.8** | 93.8 | 92.7 | 93.8 | **93.2** | 95.7 | 95.6 | 93.7 | **94.6** |

**RQ2:** The second research question investigated the effect of integrating design pattern information on the performance of state-of-the-art deep learning models. Analyzing the results of the code classification task reveals that the incorporation of design pattern information led to a more significant performance improvement than the addition of bug information alone. The results suggest that the incorporation of more static analysis information, specifically design pattern information, is beneficial for the tasks. This is particularly relevant for the code clone detection task, which relies on the BCB dataset. This dataset places a greater emphasis on the code structure and the functional role of code snippets. After adding the design pattern information, the effect improved, and the best results were obtained when combining the bug information and the design pattern information simultaneously. Our findings suggest that integrating a diverse range of analytical information can significantly enhance model performance.

## 4 Limitations and Future Work

In this study, we demonstrate the benefits of incorporating static analysis information, such as bug information and design patterns, into raw datasets to augment the performance of code-related tasks. Using the ASTNN model as a base, our results reveal that incorporating static analysis significantly enhances performance, particularly when utilizing design patterns, which provide insight into the functional attributes of source code.

### 4.1 Limitations

Despite the promising results, there exist several limitations in this study. First, the sheer volume of code snippets in the dataset restricted the use of all data when incorporating bug reports, with defects found in only 50% of the snippets. Second, while the BCB dataset, comprising real code snippets extracted from the SourceForge Java repository [19], offers a realistic perspective that aligns with actual software development practices, the OJ dataset employed in this study was not derived from a production environment. Third, the code snippets in these datasets are relatively brief, potentially failing to capture the complexity and scale of actual code sets found in software engineering projects. Fourth, this study predominantly revolves around the ASTNN model, which may limit the generalizability of the results. The improvements observed might be specific to the ASTNN and may not necessarily extend to other source code representation models. Future work should aim to investigate the applicability of the proposed approach with a variety of other baseline models to confirm its broader effectiveness. Overall, while this study focuses on integrating static analysis within raw datasets to enhance the performance of source code-related tasks, the limitations inherent to the dataset itself pose the primary constraint [15].

### 4.2 Future Work

In the scope of this research, we have observed that incorporating additional contextual information, such as bugs and design patterns, benefits deep learning models in comprehending source code, much like humans. Our preliminary findings provide promising evidence to support this claim. However, we believe that the type of context beneficial for enhancing a model's performance may vary depending on the task at hand, such as code summarization or clone detection. Therefore, we advocate for future research to delve into this uncharted territory and explore the impact of diverse types of contextual information on different tasks.

Further, how we encode and represent contextual information merits attention. For example, determining the most effective way to encapsulate the fact that a piece of code adheres to a particular design pattern presents an interesting research question. Hence, not only should we explore what context to include, but also how best to represent it for consumption by the model. In line with these objectives, we propose a few strategic directions for future research:

*Construct Diverse and Comprehensive Datasets:* Many existing datasets containing code-related information and labels are either inadequate in size or inaccessible. Therefore, it is crucial to collect and create more extensive, research-appropriate datasets [20]. Strategies might involve using models or techniques to rectify bugs in source code, rather than just identifying them, as inaccurate bug information could obstruct source code representation. Additionally, automated tools for code annotation could be employed to enrich the overall dataset. Leveraging a more efficient design pattern detection model could enable the construction of a larger labelled dataset for training purposes.

*Develop a Unified Experimental Platform:* To accommodate a variety of corpora, evaluation metrics, and benchmark methods, it is essential to establish a unified experimental platform [22]. This platform would facilitate the use of a consistent deep learning framework, allowing both traditional methods and current evaluation metrics to be implemented and minimizing the impact of different deep learning frameworks on method effectiveness. It would also enable a more comprehensive and equitable evaluation of newly proposed methods against classic baselines.

*Enhance Supplementary Information Quality*: The quality of additional information could be improved through several approaches, such as mining knowledge from crowdsourced platforms, analyzing related software products, and studying effective review styles [25]. It is possible to construct a code knowledge graph and integrate information into the training model.

*Assist with Other Software Engineering Tasks*: Automatic generation of annotation information can aid in various software engineering tasks [24]. For instance, analyzing code can generate corresponding annotations. Existing research results can help with other tasks, such as generating code through the analysis of method or class names [23]. Generating high-quality code annotations for each module may enhances the efficiency of defect localization methods based on code search and information retrieval.

In conclusion, we advocate for future research to delve deeper into the area of integrating additional context into source code representation for deep learning models. The type of context and its representation method could significantly influence the performance of different tasks, and we believe that exploring these variables presents a wealth of opportunities for future studies. Finally, in consideration of these objectives, we propose several strategic directions for future research, including the construction of more comprehensive datasets, the development of a unified experimental platform, and the focus on enhancing the quality of supplementary information. By pursuing these avenues, we can continue to advance understanding and improve the effectiveness of deep learning models in the field of software engineering. Our code and experimental data are available at https://github.com/Snowy0647/CR-with-static-analysis.